\documentstyle[preprint,prb,aps]{revtex}
\begin{document}
\draft
\title{Chaotic regimes of antiferromagnetic resonance in a 
quasi-two-dimensional easy-axis antiferromagnet 
$({\rm NH}_3)_2({\rm CH}_2)_4{\rm MnCl}_4$}
\author{M. M. Bogdan, M. I. Kobets, and E. N. Khats'ko}
\address{B. Verkin Institute for Low Temperature Physics and Engineering,\\ 
National Academy of Sciences of the Ukraine, 310164 Kharkov, Ukraine\\
E-mail: bogdan@ilt.kharkov.ua}
\date{Submitted October 23, 1998}
\maketitle
\begin{abstract}
Chaotic regimes of the microwave energy absorption are experimentally observed and
analyzed for two-dimensional metallorganic antiferromagnet $({\rm NH}_3)_2({\rm
CH}_2)_4{\rm MnCl}_4$ at low temperatures under the conditions of nonlinear
antiferromagnetic resonance. Relaxation oscillations of energy absorption are investigated in
detail. Their frequency spectra, frequency--amplitude characteristics, and dependences
of absorbed power on driving power and static magnetic field are studied. It is shown
that the dynamics of relaxation oscillations undergoes a transition to chaos by ``irregular
periods''. Peculiarities of the transition are described consistently. Among other things, the
conditions for the emergence of energy absorption regimes with a spike-like and a saw-tooth
signal structure are determined, and the characteristics of chaotic oscillations such as the
dimensions of strange attractors are calculated. The chaotic dynamics is found to be
high-dimensional with a large contribution from noise which is of deterministic origin in the
antiferromagnet under investigation.
\end{abstract}
\pacs{Fiz. Nizk. Temp. {\bf 25}, 263--276 (March 1999)}
\narrowtext

\section*{Introduction}

Chaotic resonant phenomena in magnets have become an object of intense experimental
studies in the last decade.\cite{1,2,3,4,5,6,7,8,9,10,11,12,13,14,15,16,17,18,19} These
investigations were stimulated by the progress made in the mathematical theory of chaos
predicting the universal character of chaotic phenomena irrespective of the nature of the physical
object being studied and demonstrating a nonlinear behavior.\cite{20,21,22}

Magnetic compounds possessing the properties required for the emergence of nonlinear
oscillations include first of all the crystals exhibiting an extremely weak relaxation of spin excitations.
Yttrium--iron garnet (YIG) with a low threshold for a parametric excitation of spin waves
even at room temperature has been studied most thoroughly.\cite{1,5,8,9,11,12,13} Since YIG behaves as a
ferromagnet in the magnetic respect, nonlinear chaotic effects were studied, as a rule, under
the conditions of ferromagnetic resonance (FMR) in a transverse as well as longitudinal
driving fields.

The total number of investigated nonlinear magnet is not large, and some of them exhibit
nonlinear properties only at low temperatures of the order of a few kelvins, at which phonons
are frozen out, and their interactions with magnons becomes very weak.

The effective dimensionality of crystals plays an important role for the suppression of
relaxation processes. Stepanov et al.\cite{14,15,16,17,18} investigated the class of
metallorganic compounds that are quasi-two-dimensional ferro- and aniferromagnets in the
magnetic respect. It was found that low-dimensional magnets at low temperatures go over
to a state with an anomalously low spin--lattice relaxation virtually with a threshold,
which makes it possible to excite spin waves parametrically at microwave pumping power
of the order of a few milliwatts.\cite{18} Among other things, it was found that in addition
to YIG, nonlinear ferromagnetic crystals include metallorganic compounds with a structure
similar to the $({\rm CH}_3{\rm NH}_3)_2{\rm CuCl}_4$ crystal.\cite{19} On the other
hand, it was established that chaotic oscillations are generated in the crystals 
\mbox{CuCl$_2$2H$_2$O,\cite{2} CsMnF$_3$,\cite{6,10} 
(CH$_2$NH$_3$)$_2$CuCl$_4$\cite{9}} under the conditions of antiferromagnetic 
resonance (AFMR).

The range of nonlinear effects that have been discovered and thoroughly investigated in ferro-
and antiferromagnets is quite large. These include spin-wave instabilities (Suhl instabilities
of the first- and second order),\cite{23} auto-oscillations of absorbed microwave
power,\cite{24,25} and the observation of three known scenarios of a transition to chaos:
by period doubling (Feigenbaum scenario), quasiperiodicity, and 
intermittency.\cite{20,21,22}                    

Apart from the interpretation of these nonlinear effects and the determination of the
conditions for their observation, it was found that real magnetic crystals can demonstrate a
more complex pattern of transition to chaotic regimes in resonance experiments. None of the
known scenarios is realized in pure form in such cases,\cite{26} and we must consider new
mechanisms of chaotization.\cite{3,5,27}

Hartwik et al.\cite{28} were the first to discover long ago the so-called relaxation chaotic
oscillations of microwave power in YIG with which a new scenario of a transition to chaos
by irregular periods has been associated in last decade.\cite{3,8,9,29} This effect
lies in the emergence, instead of purely periodic auto-oscillations, of irregular chaotic bursts
of absorbed power in the form of spike-like peaks or pulses with a steep leading front and
relaxing rear front under certain conditions of magnetic resonance upon an increase in the
pumping power. Theoretical approaches to the description of such oscillations and
mechanisms of their formation were made in Refs.~\onlinecite{9,29,30}, but a systematic
analysis of temporal series of experimental signals as well as of the results of numerical
simulation of relaxation oscillations was carried out only recently.\cite{31,32}

In this work, we study experimentally the regimes of chaotic behavior of the microwave
power absorbed in a two-dimensional easy-axis antiferromagnet $({\rm NH}_3)_2({\rm
CH}_2)_4{\rm MnCl}_4$ under the AFMR conditions. This compound is a typical
representative of the family of layered Heisenberg antiferromagnets 
$\mbox{[NH$_3$-(CH$_2$)$_m$-NH$_3$]MnCl$_4$ (2CmMn)}$
studied by Stepanov et al.\cite{14,15,16,17,18} The structure of these metallorganic crystals
is formed by almost quadratic layers of magnetic ions in the octahedral environment of
chlorine ions between which long chains of alkylene--ammonia molecules are located. The
small value of interlayer exchange associated with a large separation between the spins of
adjacent layers leads to a quasi-two-dimensional behavior of these systems. At temperature
$T_N = 42.6\,{\rm K}$,\cite{33} the compound 2C4Mn is transformed into the
antiferromagnetic states with the easy magnetization axis directed at right angles to the planes
of the layers. A detailed analysis of linear antiferromagnetic resonance in 2C4Mn
revealed\cite{16} that this compound has a four-sublattice noncollinear antiferromagnetic structure
with a weak ferromagnetic moment. The antiferromagnetism vector of each layer is deflected
successively from the normal to the plane of a layer through an angle of $\pm 16^{\circ}$
so that the total vector is perpendicular to the layer, and the weak antiferromagnetism vector
lies in the layer. According to estimates, the strength of interaction between the layers
(26~Oe) is extremely small as compared to both the intralayer exchange ($2H_e \simeq
1360\,{\rm kOe}$) and the intralayer uniaxial anisotropy ($H_a \simeq 0.8\,{\rm kOe}$) so
that 2C4Mn can be regarded as an almost two-dimensional antiferromagnet.

At low temperatures (of the order of a few kelvins), the related compound 
$\mbox{(NH$_3$C$_2$H$_5$)$_2$MnCl$_4$ (1C2Mn)}$ 
revealed a number of interesting nonlinear phenomena in the behavior of the
absorbed microwave power, e.g., the emergence of auto-oscillations and 
chaos.\cite{18} Here we carry out a systematic analysis of chaotic regimes of
antiferromagnetic resonance in a 2C4Mn crystal. For a driving power below 5~mW at a
temperature below 2.18~K, we observed a nonlinear absorption of the microwave field and
the emergence of relaxation oscillations with typical (extremely low) average frequencies of
the order of a few hertz. These oscillations were recorded and analyzed as temporal series
of data with the help of an analog-digital device and computer programs, which made it
possible to describe in detail description of the scenario of a transition to chaos by
irregular periods. As a result, we have analyzed qualitative changes in the behavior of
temporal series  of absorbed power as a function of variation of the parameters of static and
varying magnetic fields and carried out the Fourier analysis, obtained the spectra of
oscillations, studied the structure of strange attractors of chaotic regimes, and calculated the
quantities characterizing chaotic dynamics, i.e., dependences of frequencies of 
auto-oscillations oscillations on the driving power (in particular, we 
determined their period doubling threshold) and the dimension of chaotic 
attractors, and discussed the origin and role of noise in relaxation 
oscillations as well as a possible theoretical model for describing
relaxation oscillations in two-dimensional antiferromagnets.

\section*{Experimental technique}

Single crystals of $({\rm NH}_3)_2({\rm CH}_2)_4{\rm MnCl}_4$ were grown at room
temperature from a saturated aqueous solution in the form of thin rectangular plates with
clearly manifested lateral faces and with a typical size $5 \times 5 \times 0.3\,{\rm mm}$.
The compound $({\rm NH}_3)_2({\rm CH}_2)_4{\rm MnCl}_4$ possesses a monoclinic
symmetry of crystal lattice with the space group $P2_{1}/b$.\cite{34} Organic chains 
of ${\rm NH}_3({\rm CH}_2)_4{\rm NH}_3$ separate two-dimensional almost square layers of
octahedra Mn-${\rm Cl}_6$. The unit cell parameters are 
\mbox{$a=10.77$ \AA , $b=7.177$ \AA , $c=7.307$ \AA}.
Experiments were carried out on a reflection spectrometer with a pumping frequency
70.39~GHz at a temperature below 2.18~K. We used a cylindrical resonator with the $Q$-
factor $\sim 1000$. The sample was placed in the resonator region with predominant parallel
polarization of external static and rf fields $H \parallel h$.

In resonance experiments, the field is usually applies along the easy axis of the crystal. With
such an orientation, the splitting of AFMR branches follows the law
\begin{equation}
\label{1}\omega _{\pm }=\gamma \{\sqrt{(2H_E+H_A)H_A}\pm H\}
\end{equation}

For the frequency mentioned above, the resonance conditions are satisfied for the lower
frequency branch $\omega_-$, which was observed in our experiments. The maximum power
of the source was  5~mW. The applied magnetic field was scanned along the contour of the
AFMR line, and the driving power was varied from 0 to $-20\,{\rm dB}$. The magnetic
field orientation relative to the anisotropy axis and equilibrium directions of
antiferromagnetism vectors in adjacent planes also varied. It was found that the most intense
absorption corresponds to the symmetric orientation of the field along the crystallographic
axis $b$.

In all experiments, low-frequency modulation of electromagnetic field of frequency 50~Hz
was observed. This frequency had to play the role of the reference frequency in our
experiments. It was found later that these oscillations participated in all nonlinear processes,
and the emergence of their higher harmonics was regarded as a natural criterion of the
emergence of nonlinearity in resonance effects. The reflected signal after detection in an
analog-digital device PC ADDA-14 with a 14-bite resolution was transformed into a
computer data file. These temporal series were subsequently analyzed by using the standard
and original packets of programs created for a quantitative analysis of chaotic phenomena.

\section*{Discussion of main results}

A typical form of resonant curves for the antiferromagnet 2C4Mn at $T = 1.8\,{\rm K}$
are shown in Fig.~\ref{fig1}. For a low (less than $-15\,{\rm dB}$) microwave field power,
a typical pattern from a linear AFMR is observed, i.e., two lines from two centers
(neighboring planes). The separation between the peaks on the resonant curve can vary
depending on the orientation of the static magnetic field, and the lines can coincide when the
field is directed along the crystallographic axis $b$. For a power exceeding $-15\,{\rm dB}$,
free relaxation oscillations are generated in the range of external magnetic fields near the
peak of the high-field line above as well as below the resonant field $H = 9.34\,{\rm kOe}$
(this region is shown by the bold line on the upper curve in Fig.~1). The amplitude of these
oscillations increases with power until they become chaotic. We shall consider this regime
in detail later, and now we pay attention to another effect associated with instability of
resonance at high pumping levels. As the driving power increase above $-5\,{\rm dB}$ (see
two lower curves in Fig.~1), a jump and a discontinuity of both resonant lines are observed
with a considerable hysteresis in the static magnetic field as we move towards higher and
lower fields respectively.

This phenomenon is well known in the theory of nonlinear resonance and is associated with
the dependence of the frequency of nonlinear oscillations on their amplitude. In magnets, this
effect is manifested in that the resonant curve must become asymmetric and multiple-valued
for a pumping field $h$ exceeding the critical value, i.e., the peak must be inclined towards
lower or higher fields depending on the type of interaction between magnons. In actual
experiments, instability is observed upon a change in the static magnetic field, and the
resonant curve experiences a discontinuity or a jump. This effect was observed for the first
time in disks of yttrium-iron garnet single crystals by Weiss.\cite{35}

It follows from Fig.~1 that the jump is observed in strong fields in an increasing field,
while discontinuity takes place in weaker fields in a decreasing field. With increasing power,
the hysteresis loop increases, and the steepness of lines decreases (the scales of conditional
units for absorbed power in Fig.~1 are different for the three resonant curves: it decreases
with increasing amplitude of pumping).

It was noted above that oscillations of observed power on the segment of the resonant curve
near 9.34~kOe appear even at very low driving powers of the order of $-15\,{\rm dB}$.
The criterion of a transition to the nonlinear regime is the emergence of the second
harmonic peak in reference oscillations with frequency 100~Hz.  At the point of maximum
on the resonant curve, this peak exceeds the background noise for a power $P > -15\,{\rm
dB}$, and first spike-like peaks of absorbed power appear at the same instant.

In the case of a resonant curve with spaced peaks, an increase in the driving power induces
relaxation oscillations in the vicinity of the second peak also. As the peaks converge, the
mutual effect of the centers increases, which is noticeably reflected in the form of oscillations
of absorbed power. In the cases of closely spaced peaks for the value of the field $H_m =
8.37\,{\rm kOe}$ corresponding to a local minimum at the center of the resonant curve,
relaxation oscillations become irregular even for a low driving power.  As the power
increases, the oscillations become more and more chaotic. The time dependence of the signal
typical of the entire series of these measurements and its spectrum for the maximum value
of power are shown in Fig.~\ref{fig2}$a$.

In order to find out whether such a dependence is a consequence of additive or dynamic
noise, stochastic process, or is due to a determinate chaos, we varied in the experiments
the orientation of magnetic field and its magnitude. The variations affected strongly the type
of oscillations.

It was found that oscillations become less chaotic for the minimum deviation of the
magnitude of magnetic field from the extremal value $H_m$. By way of an example,
Fig.~\ref{fig2}$b$ shows the time behavior and spectrum of oscillations for $H =
8.4\,{\rm kOe}$ and the maximum driving power.

It also turned out that the degree of chaotization of a signal decreases considerably, and
its shape changes qualitatively when the magnetic field is directed along the crystallographic
axis $b$, when the resonant lines from two centers coincide. In this case, the characteristic
pattern of the emergence and transformation of relaxation oscillations upon a change in the
driving power is of the form shown in Fig.~\ref{fig3} for $H = 8.4\,{\rm kOe}$ (small
deviation of the field from the resonant value) and the power $P$ varying from $-10$ to
0~dB. First spikes of absorbed power appear against the background of almost linear
oscillations of frequency 50~Hz. For small pumping amplitudes, the frequency
corresponding to the emergence of spike-like peaks is low, and the intervals between them
are quite large and vary with an obvious periodicity. For a driving power of the order of $-
6\,{\rm dB}$, the signal has the form of a periodic structure of spike-like closely spaced
peaks. As the pumping amplitude increases further to power values of the order of $-3\,{\rm
dB}$, the frequency corresponding to the emergence of spike-like peaks changes
insignificantly, and subsequently decreases rapidly and becomes virtually equal to half the
previous value. In this region, the shape of the signal changes qualitatively from the spike-
like to the saw-tooth, i.e., the change in the regime of chaotic oscillations takes place.
Figure~\ref{fig4} shows the dependence of the fundamental frequency of these oscillations
on the driving power in the range from -6 to 0~dB (solid circles). The doubling
of the period of relaxation oscillations can be seen clearly in the figure. In order to plot this
dependence, we analyzed the spectra of oscillations for fixed pumping levels. It should be
noted that doubling of this period does not indicate the emergence of subharmonics of the
fundamental frequency as is usually the case in the Feigenbaum scenario, and corresponds to
the change from one oscillatory mode to another mode, their fundamental frequencies differing
by a factor of two. It was proved that this effect is preserved for other values of magnetic
field which naturally affects the values of frequencies themselves. It should also be noted that
apart from the main peaks and multiple harmonics, all spectra contain a large contribution
from noise responsible for the ``grass-like'' continuous spectrum. In the subsequent analysis,
we shall analyze in detail the dynamic and spectral structure of these oscillations and the
origin of their stochastic form.

When the power changes in the opposite direction, i.e., the amplitude of pumping decreases
(open circles in Fig.~\ref{fig4}), the frequency--amplitude dependence exhibits a hysteresis
with a displacement of the region of period doubling towards lower powers (saw-tooth pulses
exist up to $-4.5\,{\rm dB}$). The existence of essentially chaotic modes near a certain fixed
values of power, in particular upon an increase in the driving power for $P = -1, -2.25$,
and $-2.75\,{\rm dB}$ is an interesting feature of the observed transient process.

For this reason, it was natural to analyze oscillations for these selected pumping levels, but
in a wide range of applied magnetic field near the resonance point. We chose the pumping
level of $-1\,{\rm dB}$ and studied the variation of the shape of the absorbed power signal
and its spectrum upon a change in the static magnetic field within a few ten oersted near the
resonant value $H_r = 8.37\,{\rm kOe}$. The direction of the field was maintained along
the crystallographic axis $b$.

The corresponding results are presented in Fig.~\ref{fig5}. It should be noted that relaxation
oscillations occur against the background of a considerable average absorbed power that must
make a contribution to the frequency spectrum in the form of a large central peak at zero
frequency. In all calculations of the spectra analyzed here, this average value was subtracted,
and hence the given huge contribution to the central peak is absent, which allows us to see
the detailed structure of relaxation oscillations proper. It should also be noted that frequency
spectra are given in the form of frequency dependences of the amplitude of the Fourier
transform of the signal and not as logarithmic spectra of power in order to improve
detailization.

Another feature in common of all the spectra considered below is the presence of oscillations
of frequency $\nu_0 = 50\,{\rm Hz}$. These low-frequency oscillations were present as a
source of reference frequency, but they became involved in free oscillations in view of the
nonlinearity of the medium. This follows from the presence of second harmonic with
frequency $\nu = 100\,{\rm Hz}$ and the peaks that are algebraic sums of frequencies
of fundamental harmonics of relaxation oscillations and the reference frequency.

Far away from the resonant field, the absorbed power is virtually constant if we disregard
extremely low background noise in which, however, oscillations with frequency $\nu_0 =
50\,{\rm Hz}$ were always manifested (in the frequency spectrum) in our measurements. As
the field approaches the resonant level, these small-amplitude oscillations become weakly
nonlinear (second harmonic appears in the spectrum), and nearly periodic spike-like peaks
of absorbed power corresponding to peaks of the order of a few hertz in the frequency
spectrum and clearly distinguishable against the ``grass-like'' background noise appear almost
simultaneously.

A typical example of such a behavior of absorbed power is shown in Fig.~\ref{fig5}$a$ for
the field value $H = 8.51\,{\rm kOe}$. It can be seen that periodic relaxation oscillations
with a spike-like structure have been formed completely.  Small anharmonic modulation of
peak amplitudes is manifested in the frequency spectrum in the form of higher harmonics of
the fundamental frequency. All the remaining peaks can be identified as algebraic sums of
these harmonics and frequencies $\nu_0$ and $2\nu_0$.

As we approach the resonant field further, the shape of absorbed power peaks experiences
rapid qualitative changes. Figure~5$b$ shows the result of transformation of spike-like
signals into typical saw-tooth temporal series for $H = 8.47\,{\rm kOe}$. In addition to the
increase in the amplitude and relative height of frequency peaks, the emergence of linearly
increasing and decreasing segments on the time dependence of absorbed power is also worth
noting. It is remarkable that such oscillations are almost indistinguishable from classical
relaxation oscillations that are frequently encountered in electrical engineering.

A subsequent decrease in the field leads to the tendency to the formation of periodic
rectangular pulses of absorbed power. Signals of such a shape are shown in Fig.~5$c$ for
$H = 8.44\,{\rm kOe}$. It should be noted that the amplitude of oscillations does not
increase any longer, while the periodicity is enhanced, which is manifested in the frequency
spectrum.

Relaxation oscillations become completely chaotic for field values close to resonance.
Figure~5$d$ shows the corresponding temporal series of absorbed power and a typical
``grass-like'' frequency spectrum for $H = 8.39\,{\rm kOe}$. It can be seen that the
amplitudes of oscillations are much smaller than those in Fig.~5$c$, and the frequency
distribution of oscillations has become almost continuous with a sharp decrease in the
maximum peak heights to the amplitude of the 50-Hz peak of the fundamental harmonic.

As the field decreases further from the resonant value, relaxation oscillations again acquire
the spike-like shape, being essentially nonlinear. Figure~5$e$ shows for $H = 8.34\,{\rm kOe}$ the temporal series for
such anharmonic oscillations of absorbed power and their frequency spectrum with clearly
manifested peaks of multiple harmonics. A distinguishing feature of these oscillations is that
their fundamental frequency is almost half the frequency of similar spike-like oscillations
presented in Fig.~5$a$.The frequency of relaxation oscillations in general decreases as the
field decreases to the resonant value, and starts increasing after the passage of the resonance
peak.

In the magnetic field scanning in the opposite direction (i.e., upon its increase), the regimes
described above appear in the reverse order, but a hysteresis loop takes place in complete
accord with the picture shown in Fig.~1.

It was mentioned above that the selection of other values of power (for example, the
maximum power $P = 0\,{\rm dB}$) followed by scanning in the static magnetic field
results in chaotic nonlinear oscillations whose frequency structure contains higher harmonics
of the fundamental frequency as well as subharmonics against the background of a high-intensity 
continuous noise spectrum (see Figs.~2$a$ and $b$).

A comparison of temporal series also leads to the conclusion concerning clearly manifested
temperature dependence of the degree of stochastization of oscillations. The higher the
temperature, the higher the noise level in the oscillatory spectra and the extent of their
nonregularity, and vice versa. At low temperatures, we could observe relaxation oscillations
in the form of nearly rectangular pulses (such a mode was realized for $P = 5\,{\rm mW},
H = 8.3\,{\rm kOe}$, and $T = 1.7\,{\rm K}$).

Another interesting feature is the observation of the regime of an abrupt and virtually
complete disappearance of free oscillations with simultaneous doubling of the period of
nonlinear reference oscillations and the emergence of their subharmonic at the frequency
25~Hz. We can try to explain the latter effect from the point of view of the theory of chaos
control and the emergence of higher (multiple) resonances. However, we shall not consider
this problem here and analyze the structure of chaotic attractors of relaxation oscillations.

\section*{Analysis of experimental results}

The method of a nonlinear analysis of experimental temporal series has been worked out
intensely during the last decade and is described in detail in a number of reviews and
monographs.\cite{19,36,37,38} We shall use this method which involves the determination
of the linear autocorrelation function for temporal series, the determination of 
``time delay'', the construction of phase portraits of attractors in the 
corresponding ``time delay'' coordinates, the construction of interspike intervals 
and their analysis, the computation of the correlation dimension of attractors, the
determination of the noise contribution to temporal series, the source of the noise and 
possibilities of its reduction and the discussion of theoretical models of the 
observed chaotic oscillations.

The temporal series is a discrete set of values of the physical quantity (the absorbed power
$V(t_n)$ in our case), measured in equal intervals of time. A traditional characteristic of
temporal series of signals is the linear autocorrelation function\cite{37}

\begin{equation}
\label{2}C_L(\tau )=\frac{\frac 1N\sum\limits_{m=1}^N[s(m+\tau )-%
\bar s][s(m)-\bar s]}{\frac 1N\sum\limits_{m=1}^N[s(m)-\bar s]^2} ,
\end{equation}
where the average value of the signal $s(m)$ is defined in the standard manner:
\mbox{$\bar s=\frac 1N\sum\limits_{m=1}^Ns(m)$}.

Since we usually subtract the average value of series from the initial series in an analysis of
spectra, we calculated autocorrelation function for time dependences presented in Fig.~3
for modified series $\bar{s} = 0$. As a function of $\tau$, it exhibits qualitatively identical
behavior for all values of power: this is an oscillating function with a slowly decreasing
amplitude. The period of these oscillations coincides with the fundamental period of
oscillations of the signal being measured. In an analysis of nonlinear signals, autocorrelation
function is also useful for estimating ``time delay''. It is chosen\cite{19,37} equal to the
value of $\tau$ for which the autocorrelation function vanishes for the first time. In our
measurements, this time delay is approximately equal to a quarter of the fundamental period
of observed oscillations.

Figure~\ref{fig6} shows the dependence of time $\tau$ on the driving power $P$. (The unit
of measurements of $\tau$ is the principal interval $\Delta t = 4.9\,{\rm ms}$ of our
temporal series.) It can be seen that this dependence obviously correlates with the dependence
of the frequency of oscillations on the driving power presented in Fig.~4 and confirms the
existence of a threshold transition from one regime of chaotic oscillations to another. It
should be noted that the period of the correlation function corresponds to oscillations of
frequency $50\,{\rm Hz}$ for low powers and to the fundamental period of saw-tooth
oscillations for the maximum power.

The obtained value of $\tau$ can now be used for plotting phase portraits of nonlinear
oscillations. For this purpose, we shift the temporal series by $\tau$ and plot the dependence
of $V(t_n + \tau)$ on $V(t_n)$. These functions are just the time-delay coordinates.
For the temporal series corresponding to the maximum power in Fig.~3, the phase portrait
is shown in Fig.~\ref{fig7}$a$ (the value of $\tau$ is chosen equal to $49\,{\rm ms}$, and
the average value of absorbed power is subtracted from the given series). It can be seen that
the process is periodic on the whole and occurs in several stages with their own characteristic
times. In order to obtain a more detailed concept of the attractor structure, we constructed
a $1D$ map from the sequence of minimum values of the Poincar\`{e} sections of the
given attractor. These values were determined as negative values of $V(t_m + \tau)$ taken
at instants $t_m$ for which $V(t_m) = 0$. This dependence is shown in Fig.~7$b$ and
demonstrates the existence of the internal structure of the attractor and an obviously large
contribution of noise.

A detailed analysis of the time dependence $V(t)$ indicates that the noise contribution is not
additive. Indeed, irregular amplitude jumps as well as periodic oscillations of frequency
$50\,{\rm Hz}$ have different values at different stages of variation of the function $V(t)$.
This indicates a nonlinear enhancement of both factors and their participation in the chaotic
process.

In order to describe the chaotic behavior of relaxation oscillations and the effect of noise on
them quantitatively, we consider a sequence of time intervals between adjacent peaks of the
signal and the sequence of maximum values of peak amplitudes as characteristics of this
process.

These dependences of amplitude peaks and interspike intervals for the series under
investigation are compared with the relevant sequences for the series shown in Fig.~5$a$.
It can be seen that the amplitude peaks of oscillations of the absorbed power
(Figs.~\ref{fig8}$a$ and $b$) behave quite chaotically in the vicinity of a resonance and far
away from it. At the same time, the interspike intervals (Figs.~8$c$ and $d$) exhibit
a clearly manifested tendency to a quasiperiodic mode far away from the resonance
(Fig.~8$d$), but random forces acting on the system result in the stochastization of
oscillations, which is accompanied by chaotic jumps in the period of oscillations between its
four principal values.

We can try to determine whether a noise is stochastic or dynamic by calculating the
correlation dimension $D$ of the attractor under investigation. This quantity is defined
through the pair correlation integral:

\begin{equation}
\label{3}C_m(r)=\frac 2{N(N-1)}\sum\limits_{i\neq j}^N\theta (r-%
\mid \vec y_m(j)-\vec y_m(i)\mid ) ,
\end{equation}
where $N$ is the number of measurements, $r$ the correlation radius, $\vec y_m(i)$ 
the vector of dimensions $m$ in the embedding space, whose coordinates are 
\mbox{$\{V(t_{i),}V(t_i+\tau ),...V(t_i+(m-1)\tau )\}$},
and $\theta(r)$ is the theta function. This function in fact determines the number of pairs of
vectors in the $m$-dimensional space, the separation between which is smaller than the preset
distance $r$. While determining the distance, we presume that the cells into which the phase
space is divided have the cubic shape.\cite{39} The dimension $D$ is the limit of the
expression
\begin{equation}
\label{4}D=\lim \limits_{m\rightarrow \infty }\frac{d\ln C_m(r)}{\ln (r)}
\end{equation}
and is usually calculated on the interval $r$ in which the values of the correlation function
are not very small.

The sequence of calculated correlation sums for the initial temporal series corresponding
to the maximum power in Fig.~3 is presented in Fig.~\ref{fig9} on logarithmic scale (the
curves correspond to the variation of $m$ from 1 to 9 from top to bottom). Numerical
differentiation reveals a flat segment according to which the dimension of a strange attractor
can be estimated. It was found that it is slightly larger than two ($2.25 \pm 0.1$), but the
strong effect of dynamic noise following from the characteristic increase in the steepness of
the curves for $\ln(r) < 2.5$ does not allow us to establish the existence of exact limit. We
are inclined to interpret the latter quantity as the dimension of a regular attractor. On the
other hand, the slope of the curves in the region $1 < \ln(r) < 2.5$ also demonstrates the
tendency to a limit that can be estimated as $4.9 \pm 0.1$. Such a limiting value can be
regarded as the total dimension of the attractor, containing the contribution from a
regular attractor and a deterministic noise. A slight increase in the dimension for large
values of $m$ for small $r$ is associated with the contribution of ``white'' instrumental
noise. The above comparative analysis of functional dependences of temporal series and their
spectra also confirms this conclusion. Thus, the analysis of the correlation dimension
leads to the conclusion concerning the dynamic nature of noise in the relaxation oscillations
under investigation and indicates a multidimensional chaotic dynamics and, generally
speaking, multimode excitations in the resonance system in question. The extent of its
stochastization is quite high, which follows from the estimate of correlation dimension
in the range of small $r$. It should be noted that recent investigations of parametric
resonance in a related metallorganic antiferromagnet\cite{31} also confirms the deterministic
origin of noise in these compounds.

The existing theories of relaxation oscillations\cite{3,8,9,29,32} make it possible to describe
the emergence of high-dimensional chaos on the basis of multimode models. The main
idea behind the mechanism of emergence of relaxation oscillations can be demonstrated even
by using a two-mode model in which it is assumed that the resonance excitation conditions
are satisfied for one mode and are not observed for the other mode. Under the action of
pumping, such a pair of coupled nonlinear oscillators reproduces quantitatively the behavior
of relaxation oscillations. A quantitative theory of this effect for antiferromagnets has not
been developed as yet. However, a theoretical description of this phenomenon in the
approximation of two spins simulating the sublattices subjected to resonant transverse and
longitudinal pumping appears as promising. Consequently, chaotic relaxation oscillations can
be described qualitatively as the dynamics of a nonlinear oscillator under resonance
conditions, but under the action of certain random forces (the inclusion of the effect of the
second oscillator). Such a system may have at least two stable states the transition between
which can lead to the emergence of spike-like and saw-tooth time dependences of absorbed
power. The features and diversity of existing chaotic modes are obviously determined by the
time of residence of the system in the equilibrium states and the rate of transient processes.
Such a system can obviously have a high degree of stochastization, an attempt to create
regular attractors in it will lead to regimes in which such attractors coexist with a well-
developed dynamic noise. However, a quantitative theory of such chaotic oscillations should
apparently be constructed on the basis of a multimode model according to the numerical
analysis carried out by Moser et al.\cite{32}

The theorem on dimension that has been formulated recently for systems with a dynamic
noise indicates in its simplified formulation the additivity of the dimension of a regular
attractor and a noise.\cite{40} In this sense, these can be separated, and the question of
elimination of noise from a signal, at least the noise reduction, and isolation of a
regular signal from the data on temporal series appears as justified.

There are effective methods of suppression of noise effect on the useful
signal.\cite{38} These methods are extremely effective for reduction of external additive
noise, but their application in the case of a dynamic noise should be verified in each specific
case. The algorithm of purification of a signal in the simplest form can be described as
follows. In the chosen embedding space whose dimension is larger than the sum of the
predicted dimension of the regular attractor and dynamic noise, the nearest neighbors of the
preferred vector of state are selected, and its central coordinate is averaged over the values
of relevant coordinates of the found neighbors. The obtained sequence of new data is the
result of one iteration that can be repeated. Such an algorithm can be optimized as well as
the choice of required parameters (correlation radius, dimension, etc.; see
Ref.~\onlinecite{38}). Such an algorithm will be used below for analyzing the chaotic
temporal series measured by us.

On the other hand, the above algorithm in the simplest form includes the conventional
method of data averaging over nearest and next to nearest neighbors in the series. In this
case, the dimension of the embedding space is equal to unity, and the number of
neighbors is fixed. It can easily be verified that in spite of its very simple form, the
procedure operates as a high-frequency filter and does not change the complex low-frequency
spectrum of chaotic oscillations. We shall apply this procedure also to analyze the results.

We chose the object of investigation in the form of a chaotic attractor obtained from the
temporal series presented in Fig.~2$b$. The results of analysis are shown in
Fig.~\ref{fig10} (it should be noted that the average value is subtracted from the terms of
the series). Figures~10$a$ and $b$ show the phase portrait on the plane $\{V(t), V(t +
\tau)\}$, where $\tau = 29.5\,{\rm ms}$, and simultaneously the map for the data on
the Poincar\`{e} cross section (see above), while Figs.~10$c, d, e$, and $f$ contain the
dependences for the data ``corrected'' by the method of averaging and the optimized method
of noise reduction described above respectively (averaging was carried out twice over five
points, and  tenfold iterations were used in the optimized method).

Figures~10$e$ and $f$ clarify the internal structure of a regular attractor. After the effect
of noise becomes weaker, its phase portrait resembles a strange multiband attractor. The
analysis of correlation dimension makes it possible to characterize quantitatively both the
regular attractor as well as the residual contribution of deterministic noise. The results of
analysis are presented in Fig.~\ref{fig11}. It should be noted that prior to calculation
correlation functions for the temporal series under consideration, we initially normalized all
values to a unit interval by the formula 
\mbox{$\tilde V(t_i)=(V(t_i)-V_{\min })/(V_{\max }-V_{\min })$}, where $V_{\min}$ and
$V_{\max}$ are the minimum and maximum values of the signal in the series. It was found
that the dimensionality of a regular attractor can be estimated as $2.15
\pm 0.05$, and the total dimensionality with the contribution of deterministic noise as $3.25
\pm 0.05$. Spectral analysis of these ``improved'' results also indicate that the quantitative
contribution of noise remained quite large, and the resultant attractor possesses a high
dimensionality as before.

Thus, chaotic dynamics in the nonlinear antiferromagnetic resonance in low-dimensional
antiferromagnets is high-dimensional, the extent of stochastization of oscillations is high, and
noise has a deterministic origin and serves as a decisive factor in nonlinear dynamics of these
magnets.

Finally, we formulate the following conclusions following from our analysis.
\begin{enumerate}
\item Peculiarities of a transition to chaos by ``irregular periods'' in a $2D$
metallorganic antiferromagnet with an ``easy axis'' type anisotropy are experimentally
observed and studied in detail under conditions of nonlinear antiferromagnetic resonance.
\item It is shown that relaxation oscillations of absorbed power are generated for very low
energy levels of microwave field and have a low frequency of fundamental harmonic (of the
order of a few hertz). No multiple harmonics are observed experimentally at kilohertz and
higher frequencies.
\item Relaxation oscillations at low values of driving power exist in the form of generally
periodic sequence of spike-like peaks of absorbed power. The frequency spectrum contains
components of fundamental frequency corresponding to the emergence of spikes as well as
multiple harmonics, which demonstrates a nonlinear nature of the process.
\item As the pumping amplitude increases, the phenomenon of period doubling is observed
in the time dependence of absorbed energy of microwave field. The shape of the signal is
simultaneously transformed from the spike-like to the saw-tooth type having segments with
linearly increasing and linearly decreasing absorption. An analysis of the
frequency--amplitude dependence of oscillations and their linear autocorrelation function gives
quantitative characteristics of this transition.
\item A similar effect is observed at a fixed level of pumping, but upon a change in the value
of static magnetic field near its resonant value.
\item With increasing power, relaxation oscillations become chaotic. The spectrum of such
oscillations is continuous and has a ``grass-like'' form, but the peaks of fundamental
harmonics are still distinguishable. The phase portrait of these oscillations has the form of a
strange attractor experiencing a strong influence of noise. Stochastization of oscillations,
however, is not a result of influence of an additive instrumental noise.
\item The quantitative characteristics of such a strange attractor are calculated. The one-
dimensional map corresponding to the given attractor demonstrates a tendency to regular
movement in spite of chaotic time dependence of relaxation oscillations. An analysis of
correlation dimension indicates the high-dimensional chaos dynamics and the
deterministic nature of noise in the magnetic system under investigation. The possibility of
formal separation of the regular movement and the noise contribution with the help of
nonlinear methods of noise reduction being developed is considered.
\item The applicability of the theoretical model of finite number of coupled spins under the
action of the parametric and transverse pumping to the construction of a quantitative theory
of the scenario of transition to chaos by ``irregular periods'' is discussed briefly. The
transition can be regarded as a universal phenomenon in low-dimensional ferro- and
antiferromagnets under nonlinear resonance conditions.
\end{enumerate}

\section*{Acknowledgements}

The authors are grateful to S. V. Volotskii for fruitful discussions and to H. Kantz, R.
Hegger, and Th. Schreiber for valuable advice and help in carrying out a number of
calculations.

\begin{figure}
\caption[]{Amplitude--field dependences of nonlinear antiferromagnetic resonance for a
2C4Mn crystal for different values of driving power. The bold segment on the upper curve
in the vicinity of the high-field peak denotes the range of relaxation oscillations. The
hysteresis loop observed for a power exceeding $-5\,{\rm dB}$ increases with driving
power.}
\label{fig1}
\end{figure}
\begin{figure}
\caption[]{Chaotic oscillations of absorbed power and their frequency spectrum for a driving
power of 5~mW for a resonant curve with closely spaced peaks in fields $H_m = 8.37\,{\rm
kOe}\,(a)$ and 8.4~kOe ($b$).}
\label{fig2}
\end{figure}
\begin{figure}
\caption[]{Evolution of time dependences of absorbed power for a change in the microwave
power level from $-10$ to 0~dB for a static magnetic field $H = 8.4\,{\rm kOe}$.}
\label{fig3}
\end{figure}
\begin{figure}
\caption[]{Frequency of relaxation oscillations as a function of driving power.
Solid and open circles correspond to an increase and decrease in the driving power
respectively. The threshold effect of oscillation period doubling is of the hysteresis type.}
\label{fig4}
\end{figure}
\begin{figure}
\caption[]{Temporal series of absorbed power and their spectra for different values of static
magnetic field for the driving power $P = -1\,{\rm dB}:\,(a)\, H = 8.51\,{\rm kOe}$,
relaxation oscillations with a spike-like structure and a high stochastization level; $(b)\,H =
8.47\,{\rm kOe}$, the result of transformation of spike-like signals into saw-tooth
signals; $(c)\,H = 8.44\,{\rm kOe}$, saw-tooth relaxation oscillations with linearly
increasing and decreasing segments; $(d)\,H = 8.39\,{\rm kOe}$, chaotic temporal series
with an intense ``grass-like'' frequency spectrum;  $(e)\,H = 8.34\,{\rm kOe}\,(e)$, nearly
regular anharmonic oscillations of absorbed power of a spike-like shape. All the spectra
contain the peak of the fundamental frequency of relaxation oscillations of the order of
several hertz and its higher harmonics as well as the peak of the reference frequency 50~Hz
and combination frequencies.}
\label{fig5}
\end{figure}
\begin{figure}
\caption[]{Dependence of ``time delay'' $\tau$ on the pumping power $P$. The value of
$\tau$ is measured in units of the temporal series period $\Delta t = 4.9\,{\rm ms}$.}
\label{fig6}
\end{figure}
\begin{figure}
\caption[]{Dynamic characteristics of relaxation oscillations for the driving power $P =
0\,{\rm dB}$ and $H = 8.4\,{\rm kOe}$, whose time dependence is shown in Fig.~3:
($a$) phase portrait in the time-delay coordinates ($\tau = 49\,{\rm ms}$) and ($b$) one-
dimensional map constructed from the sequence of minimum values of the Poincar\`{e}
cross section of the phase portrait taken at the instants of time $t_m$ at which $V(t_m) =
0$.}
\label{fig7}
\end{figure}
\begin{figure}
\caption[]{Amplitude peaks and interspike intervals for relaxation oscillations 
for driving power $P = 0\,{\rm dB}$ and $H = 8.4\,{\rm kOe}$ 
(see Fig.~3) ($a, c$), and $P = 1\,{\rm dB}$ and $H = 8.51\,{\rm kOe}$ 
(see Fig.~5$a$) ($b, d$).}
\label{fig8}
\end{figure}
\begin{figure}
\caption[]{Dependences of logarithms of correlation integrals on the logarithm of the
distance between vectors in the $m$-dimensional embedding space calculated from the initial
integral data for the attractor presented in Fig.~7.}
\label{fig9}
\end{figure}
\begin{figure}
\caption[]{Phase portraits and one-dimensional map for relaxation oscillations for the
driving power 5~mW in the field $H = 8.4\,{\rm kOe}$ in the case of closely spaced peaks
on the resonant curves (see Fig.~2$b$) for original data on temporal series (after the
subtraction of the average value) ($a,b$), for doubly averaged data for five nearest neighbors
($c,d$), and for data ``improved'' by the optimized method of noise suppression as a result
of ten iterations ($e, f$).}
\label{fig10}
\end{figure}
\begin{figure}
\caption[]{Dependences of logarithms of correlation integrals on the logarithm of the
distance between vectors in the $m$-dimensional embedding space calculated for data
``improved'' by the optimized method (see Figs.~10$e,f$).}
\label{fig11}
\end{figure}

\begin{references}
\bibitem{1}G. Gibson and C. Jeffries, Phys. Rev. {\bf A29}, 811 (1984).
\bibitem{2}H. Yamazaki, J. Phys. Soc. Jpn. {\bf 53}, 1155 (1984).
\bibitem{3}F. Waldner, R. Badii, D. R. Barberis, et al., J. Mag. Mag. Mater. {\bf 54--57},
1135 (1986).
\bibitem{4}H. Yamazaki and M. Warden, J. Phys. Soc. Jpn. {\bf 55}, 4477 (1986).
\bibitem{5}F. M. de Aguiar and S. M. Rezende, Phys. Rev. Lett. {\bf 56}, 1070 (1986).
\bibitem{6}A. I. Smirnov, Zh. Eksp. Teor. Fiz. {\bf 90}, 385 (1986) [Sov. Phys.--JETP {\bf
63}, 222 (1986)].
\bibitem{7}H. Yamazaki, M. Mino, H. Nagashima, and M. Warden, J. Phys. Soc. Jpn. {\bf
56}, 742 (1987).
\bibitem{8}P. Bryant, C. Jeffries, and K. Nakamura, Phys. Rev. {\bf A38}, 4223 (1988).
\bibitem{9}M. Warden and F. Waldner, J. Appl. Phys. {\bf 64}, 5386 (1988).
\bibitem{10}A. I. Smirnov, Zh. Eksp. Teor. Fiz. {\bf 94}, 185 (1988) [Sov. Phys.--JETP
{\bf 67}, 969 (1988)].
\bibitem{11}T. L. Carrol, L. M. Pecora, and F. J. Rachford, J. Appl. Phys. {\bf 64}, 5396
(1988).
\bibitem{12}H. Benner, F. Rodelsperger, H. Seitz, and G. Wiese, J. Phys. {\bf C8}, 1603
(1988).
\bibitem{13}P. E. Wigen, H. Doetsch, Y. Ming, et al., J. Appl. Phys. {\bf 63}, 4157
(1988).
\bibitem{14}A. A. Stepanov, M. I. Kobets, and A. I. Zvyagin, Fiz. Nizk. Temp. {\bf 9},
764 (1983) [Sov. J. Low Temp. Phys. {\bf 9}, 391 (1983)].
\bibitem{15}A. I. Zvyagin, M. I. Kobets, V. N. Krivoruchko, et al., Zh. Eksp. Teor. Fiz.
{\bf 89}, 2298 (1985) [Sov. Phys.--JETP {\bf 62}, 1328 (1985)].
\bibitem{16}A. A. Stepanov, V. A. Pashchenko, and M. I. Kobets, Fiz. Nizk. Temp. {\bf
14}, 550 (1988) [Sov. J. Low Temp. Phys. {\bf 14}, 304 (1988)].
\bibitem{17}A. A. Stepanov, V. A. Pashchenko, and M. I. Kobets, Fiz. Nizk. Temp. {\bf
14}, 1212 (1988) [Sov. J. Low Temp. Phys. {\bf 14}, 669 (1988)].
\bibitem{18}A. A. Stepanov, A. I. Zvyagin, S. V. Volotskii, et al., Fiz. Nizk. Temp. {\bf
15}, 100 (1989) [Sov. J. Low Temp. Phys. {\bf 15}, 57 (1989)].
\bibitem{19}H. Yamazaki and M. Mino, Progr. Theor. Phys. Suppl. No.~98, 400 (1989).
\bibitem{20}J.-P. Eckmann, Rev. Mod. Phys. {\bf 53}, 643 (1981).
\bibitem{21}M. Feigenbaum, Usp. Fiz. Nauk {\bf 141}, 343 (1983) [Sov. Phys.--Uspekhi
{\bf 26}, (1983)].
\bibitem{22}J.-P. Eckmann and D. Ruelle, Rev. Mod. Phys. {\bf 57}, 617 (1985).
\bibitem{23}V. E. Zakharov, V. S. L'vov, and S. S. Starobinets, Usp. Fiz. Nauk {\bf
114}, 609 (1974) [Sov. Phys.--Uspekhi {\bf 17}, 896 (1975)].
\bibitem{24}V. V. Zautkin, V. S. L'vov, and S. S. Starobinets, Zh. Eksp. Teor. Fiz. {\bf
63}, 182 (1973) [Sov. Phys.--JETP {\bf 36}, 96 (1973)].
\bibitem{25}V. V. Zautkin, V. S. L'vov, and S. S. Starobintes, Fiz. Tverd. Tela
(Leningrad) {\bf 16}, 678 (1974) [Sov. Phys.--Solid State {\bf 16}, 446 (1974)].
\bibitem{26}H. Benner, F. Roedelsperger, and G. Wiese, in: {\em Nonlinear Dynamics in
Solids} (ed. by H. Thomas), Springer, Berlin, Heidelberg (1992).
\bibitem{27}F. M. de Aguiar, A. Azevedo, and S. M. Rezende, Phys. Rev. {\bf B39}, 9448
(1989).
\bibitem{28}T. S. Hartwick, E. R. Peressini, and M. T. Weiss, J. Appl. Phys., Suppl. {\bf
32}, 223S (1961).
\bibitem{29}F. Waldner, D. R. Barberis, and H. Yamazaki, Phys. Rev. {\bf A31}, 420
(1985).
\bibitem{30}F. Waldner, J. Phys. {\bf C21}, 1243 (1988).
\bibitem{31}M. Warden, Phys. Rev. {\bf E48}, R639 (1993).
\bibitem{32}H. R. Moser, P. F. Meier, and F. Waldner, Phys. Rev. {\bf B47}, 217 (1993).
\bibitem{33}H. Arend, K. Tichy, K. Baberschke, and F. Rys, Solid State Commun. {\bf
18}, 999 (1976).
\bibitem{34}K. Tichy, J. Benes, R. Kind, and H. Arend, Acta Cryst. {\bf B36}, 1355
(1980).
\bibitem{35}M. T. Weiss, Phys. Rev. Lett. {\bf 1}, 239 (1958).
\bibitem{36}N. Gershenfeld, in: {\em Directions in Chaos}, vol.~2 (ed. by Hao Bai-lin),
World Scientific, Singapore (1988).
\bibitem{37}H. D. I. Abarbanel, R. Brown, J. J. Sidorowich, and L. Sh. Tsimring, Rev. Mod.
Phys. {\bf 65}, 1331 (1993).
\bibitem{38}H. Kantz and Th. Schreiber, {\em Nonlinear Time Series Analysis}, Cambridge
Univ. Press, Cambridge (UK) (1997).
\bibitem{39}I. S. Aranson, A. M. Reiman, and V. G. Shekhov, in: {\em Nonlinear Waves.
Dynamics and Evolution} (ed. by A. V. Gaponov-Grekhov and M. I. Rabinovich), [in
Russian], Nauka, Moscow (1989).
\bibitem{40}M. R. Muldoon, D. S. Broomhead, J. P. Huke, and R. Hegger, Dynamics and
Stability of Systems {\bf 13}, 175 (1998).
\end{references}
\end{document}